\documentclass{ws-ijmpd}

\def \b{\beta}
\def \l{\lambda}
\def \L{\Lambda}
\def \g{\gamma}
\def \d{\delta}
\def \k{\kappa}

\def \be{\begin{equation}}
\def \ee{\end{equation}}
\def \ben{\begin{eqnarray}}
\def \een{\end{eqnarray}}

\begin{document}

\markboth{A. S. Majumdar, N. Mukherjee}
{Braneworld Black Holes in Cosmology and Astrophysics}

%%%%%%%%%%%%%%%%%%%%% Publisher's Area please ignore %%%%%%%%%%%%%%%
%
\catchline{}{}{}{}{}
%
%%%%%%%%%%%%%%%%%%%%%%%%%%%%%%%%%%%%%%%%%%%%%%%%%%%%%%%%%%%%%%%%%%%%

\title{BRANEWORLD BLACK HOLES IN COSMOLOGY AND ASTROPHYSICS}

\author{A. S. MAJUMDAR\footnote{e-mail:archan@bose.res.in}}

\author{N. MUKHERJEE\footnote{e-mail:nupur@bose.res.in}}

\address{S. N. Bose National Centre for Basic Sciences\\
Block JD, Sector III, Salt Lake, Kolkata 700098, India}

\maketitle

\begin{history}
\received{Day Month Year}
\revised{Day Month Year}
\comby{Managing Editor}
\end{history}

\begin{abstract}

The braneworld description of our universe entails a large extra dimension
and a fundamental scale of gravity that might be 
lower by several orders of magnitude 
compared to the Planck scale. An interesting consequence of the braneworld 
scenario is in the nature of spherically symmetric vacuum solutions to
the brane gravitational field equations which could represent black holes with
properties quite distinct compared to ordinary black holes in $4$-dimensions.
We discuss certain key features of some braneworld black hole geometries.
Such black holes are likely to have  diverse cosmological and astrophysical 
ramifications. The cosmological evolution of primordial braneworld black
holes is described highlighting their longevity due to modified evaporation
and effective accretion of radiation during the early braneworld high energy
era. Observational abundance of various evaporation products of the black
holes at different eras impose  constraints on their initial mass
fraction. Surviving primordial black holes could be candidates of dark matter
present in galactic haloes. We discuss gravitational lensing by braneworld
black holes. Observables related to the relativistic images of strong field
gravitational lensing could in principle be used to distinguish between
different braneworld black hole metrics in future observations. 

\end{abstract}


\section{Introduction}

There has been widespread activity in braneworld gravity in
recent times\cite{maartens}. The braneworld scenario of our universe
opens up
the fascinating possibility of the existence of large extra spatial
dimensions by ensuring that the
standard model fields are confined to the $3$-brane, whereas gravity
could also propagate into the higher dimensional bulk. Several braneworld
models have been studied in the literature, the most popular amomg them
being the Arkani-Hamed, Dimopoulos and Dvali (ADD) model\cite{arkani}, and the
two Randall-Sundrum (RS) models\cite{randall1,randall2}. In the ADD 
model\cite{arkani} there
could be $n$ large compact extra dimensions with radius $l$, with $n\ge 2$
providing a possible resolution of the hierarchy problem of particle
physics. The RS-I model\cite{randall1} is motivated from similar
considerations and consists
of two opposite tension branes, with our universe stipulated to be the
negative tension brane.
In the  RS-II model\cite{randall2} which has been the inspiration for
an extensively developed braneworld cosmology\cite{brax}, the AdS5 bulk can
have infinite
size. Confinement of the standard model fields is achieved by a positive
tension brane, and a negative cosmological constant for the AdS5 bulk
with curvature radius $l$. The resultant modification of the Newtonian
gravitational
potential in braneworld models\cite{garrtan,giddings2} is of the order 
$1/r^3$ at distances $r \ge l$ where $l$ is the scale of the extra 
dimension(s). The failure of current experiments using torsion pendulums
and mechanical oscillators
to observe departures from
Newtonian gravity at small scales have set the
upper limit of $l$ in the sub-milimeter
region, i.e., $l \le 0.2 {\mathrm mm}$\cite{long}.

It has been recently realized that the possibility of observing signatures
of modified gravity in braneworld models with large extra dimensions exists
in particle accelerator experiments. This is due to the fact that in braneworld
gravity the fundamental $5$-dimensional Planck scale $M_5$ can be much below 
the $4$-dimensional Planck scale $M_4$. If the scale of the extra dimension is
not much below the limit obtained from table-top experiments\cite{long},
then the corresponding $M_5$ could be as high as the order of a TeV. 
Thus it is possible for mini $5$-dimensional black holes to be produced in
particle collisions with centre of mass energy of TeV order\cite{giddings}.
The production cross-sections of various higher dimensional black holes
in future accelerators such as the LHC has been studied and possible signatures
in the form of properties of end products of the Hawking evaporation from
these black holes have been enlisted\cite{dimopoulos}. High energy cosmic
ray showers could similarly produce small higher dimensional black holes,
the possible signatures from which have also been studied\cite{feng}.
It hence appears that observational signatures of braneworld gravity may
be a distinct possiblity in the near future from several avenues.

The analysis of gravitational field equations on the brane is 
conceptually complicated due to the fact that the propagation of gravity
into the bulk does not permit the treatment of the brane gravitational
field equations as a closed form system\cite{shiromizu}. This makes the
task of studying gravitational collapse on the brane rather 
difficult\cite{germani}. Garriga and
Tanaka\cite{garrtan} first incorporated the effect of the Kaluza-Klein
modes on the metric outside a spherically symmetric and static matter
distribution on the brane in the form of the $1/r^3$ correction to
the gravitational potential.  Since then though no exact solution of
the full $5$-dimensional bulk field equations have been found, various
solutions representing braneworld black holes have been obtained 
based on different
configurations for the projected $5$-dimensional Weyl tensor on the 
brane. Among them the one of the Reissner-Nordstrom type with negative
tidal charge\cite{dadhich} originating from the Weyl term has been
discussed in details in the literature. The projection onto the brane of the 
$5$-dimensional Schwarzschild solution given by the Myers-Perry 
metric\cite{myers} could effectively describe a small braneworld black
hole of size $r \le l$. The properties of such black holes have been
investigated in details\cite{argyres}. The mechanism of Hawking evaporation
has been formulated into the brane and the bulk as well\cite{kanti}, and
special features have been observed in the interaction of $5$-dimensional
black holes with branes\cite{frolov1}. 

Braneworld black holes have been the subject of recent phenomenological 
interest in the arena of cosmology. Primordial black holes have potentially
diverse ramifications on several eras of cosmological evolution.
If braneworld black holes are formed in the early universe, their effect
on subsequent dynamics could be not only varied but also widely different
from those of primordial Schwarzschild black holes in standard cosmolgy.
This is due to the reasons that braneworld black holes have different
properties compared to ordinary $4$-dimensional black holes, and also
due to the entirely modified evolution of the very early stages of the
universe in the braneworld scenario\cite{maartens,brax}. In particular,
the Hubble expansion is modified by the presence of a term proportional
to the square of the energy density on the right hand side of the
Friedmann equation, which dominates the dynamics during the very early
high energy phase. Primordial braneworld black holes that could be
produced due to the collapse of horizon-sized density perturbations, have
a lower temperature and evaporate slowly compared to the standard 
$4$-dimensional black holes as a consequence of their different 
geometry\cite{guedens}. Furthermore, the modified braneworld cosmological
evolution of the universe enables accretion from the surrounding radiation
to be effective towards increasing the mass and longevity of the black
holes\cite{majumdar2,clancy}. Primordial black holes could survive till
many different eras in such a scenario, thereby contributing to the
energy density of the universe. The Hawking evaporation products at the
end of their life-cycles could have a significant bearing on several
cosmolgical processes. Observational results such as the background gamma
ray spectrum could hence be used to impose constaints on the initial mass 
spectrum of the black holes. These could be again significantly 
modified\cite{liddle} compared to those that have been obtained for
primordial Schwarzschild black holes in standard cosmology.

The prospect of survival of primordial black holes up to present times
in the braneworld scenario naturally raises the question as to whether
they could be a significant fraction of cold dark matter in galactic 
haloes, and also what role they could have in structure formation.
In order to address these issues it is first important to obtain observational
evidences of their existence and to determine their mass ranges. The two
likely avenues for obtaining observational signatures from black holes
that may be present in our galactic halo are through gravitational waves
from coalescing black hole binaries, and through gravitational lensing
of light sources by the black holes. It has been shown that energy exchange
between neighbouring black holes that are formed in the high energy 
braneworld era facilitates the later formation of black hole binaries
through gravitational interaction\cite{majumdar3}. It has been also
argued that binaries of primordial black holes in the braneworld scenario
could emit gravitational waves observable by future detectors\cite{inoue}.
The study of gravitational lensing by braneworld black holes has been
very recently undertaken. The deflection of light propagating on the brane due
to bulk effects has been calculated\cite{frolov3}. The expressions for 
the weak field bending angle 
of light in certain braneworld metrics has been obtained\cite{kar}. Further,
the various lensing quantities for one possible braneworld black hole geometry
have been obtained and compared to those for the Schwarzschild black hole
in the weak field limit\cite{majumdar4}. The richer phenomenology of strong
field gravitational lensing such as the positions and magnifications of
relativistic images is being investigated\cite{eiroa} in several 
braneworld geometries, and the values of observational parameters computed
for a candidate lense\cite{whisker}.

The plan of this review is as follows. In the next section we describe 
some candidate geometries for braneworld black holes. The arbitrariness of
the projected bulk Weyl term on the brane is responsible for the 
existence of a number of possible solutions. We discuss some distinctive
properties of a few of them. The aim of this article is to highlight
the progress made in understanding the impact of having braneworld
black holes in cosmological and astrophysical processes. In order to make
this review a bit self-contained we provide a brief description of the
essential features of the cosmology of the braneworld scenario in
section~3. The stage is then set for a somewhat detailed analysis of the
cosmological evolution of primordial braneworld black holes in section~4. 
Here we try to emphasize the key differences from the consequences of
a population of primordial Schwarzschild black holes in standard cosmology.
We begin section~5 with a skeletal description of the theoretical framework of
gravitational lensing. We then present some recent results on braneworld
lensing quantities and observables for some of the geometries described 
earlier. The underlying spirit here is the attempt to possibly discriminate 
between different gravity and braneworld gravity models by future observations.
A summary is presented and some concluding remarks are made in section~6.

\section{Spherically symmetric and static vacuum solutions on the brane}

Obtaining the gravitational field due to a localysed matter distribution
on the brane has been an involved and challenging task right since the
inception of braneworld models. This question is the forebearer of the
problem of finding the final state of gravitational collapse on the
brane, which is of central importance concerning the existence of black hole
solutions in the braneworld scenario.
The process of gravitational collapse in the braneworld scenario is
much complicated compared to general relativity because in the former case
whereas matter is confined to the brane, the gravitational field can
also access the extra dimension.
The corrections to the Newtonian
potential of a point mass $M$ at large distances due to the extra dimension
were calculated to be\cite{randall1,garrtan} 
\be
V(r) = \frac{2M}{M_4^2 r}\biggl(1+\frac{2l^2}{3r^2}\biggr)
\label{gravpot}
\ee
The failure of current experiments to detect such corrections at
sub-milimeter scales have set the upper limit on the curvature radius
$l$ of the $5$-th dimension as $l \le 0.2 {\mathrm mm}$\cite{long}.
 
The effect of Kaluza-Klein modes on the metric exterior to
a static and spherically symmetric matter distribution on the brane
was considered by Garriga and Tanaka\cite{garrtan}.
They obtained a solution in the weak field limit given by
\be 
dS_4^2 = -\biggl(1-\frac{2M}{M_4^2r} + \frac{4Ml^2}{3M_4^2r^3}\biggr)dt^2 +
\biggl(1 + \frac{2M}{M_4^2r} +\frac{2Ml^2}{3M_4^2r^3}\biggr)(dr^2 + 
r^2 d\Omega^2)
\label{gartanmetric}
\ee
Note that this solution is quite different from the Schwarzschild metric
and that the gravitational potential obtained from this metric 
(\ref{gartanmetric}) has $1/r^3$ corrections (\ref{gravpot}) compared 
to the Newtonian potential. Further perturbative 
studies\cite{giddings2,sasaki} 
have also
established that the first weak field correction to the Newtonian
potential on the brane is proportional to $1/r^3$.

The projected Weyl term $E_{\mu\nu}$ on the brane carries the imprint
of Kaluza-Klein modes that could be relevant in the process of gravitational
collapse. If the Weyl term vanishes, then the standard Schwarzschild 
solution in $4$ dimensions can be assumed as the simplest black hole solution
on the brane by `stacking' it into the extra dimension. Such a vacuum solution
of the $4$-dimensional  Einstein equation is of the `black string' 
type\cite{chamblin}, 
and can be generalised to the case of a cosmological constant in 
$4$ dimensions as well\cite{anderson}. Subsequently, it was shown that the
 black string
is unstable to large-scale perturbations\cite{gregory}.
Another solution to the vacuum $4$-dimensional Einstein field equations is
obtained
by setting the $4$-dimensional cosmological constant to zero (as in 
Eq.(\ref{cosmconst}),
thus obtaining a relation between the brane tension and the Ads 
radius (\ref{tenserad})). Since the projected Weyl tensor on the brane is
divergence free for the vacuum case, one gets for static solutions 
a closed system of 
equations given by\cite{shiromizu}
\ben
R_{\mu\nu} &=& -E_{\mu\nu}\nonumber \\
R_{\mu}^{\mu} &=& 0\nonumber \\
\nabla^{\mu}E_{\mu\nu} &=& 0
\label{fieldsoln}
\een
Dadhich et al\cite{dadhich} have prescribed the mapping of the $4$-dimensional
general relativity solution with traceless energy momentum tensor of
the Einstein-Maxwell type to a vacuum braneworld solution in $5$ 
dimensions with
the correspondence
\be
\kappa^2 T_{\mu\nu} \leftrightarrow -E_{\mu\nu}
\label{corres}
\ee

An exact black hole solution to the
effective field equations on the brane of the Reissner-Nordstrom type 
was given with the 
above correspondence (\ref{corres}) as\cite{dadhich}
\be
dS_4^2=-\left(1-\frac{2M}{M_4^2r}+\frac{Q}{r^2}\right)dt^2\\+
\left(1-\frac{2M}{M_4^2r}+\frac{Q}{r^2}\right)^{-1}dr^2\\ +r^2(d\Omega^2)
\label{tidalmetric}
\ee
where $Q < 0$ is not the electric charge of the conventional 
Reissner-Nordstrom metric, but the negative `tidal charge' arising from the
projection on to the brane of the gravitational field in the bulk.
Since the black hole mass $M$ is the source of the bulk Weyl field,
the tidal charge $Q$ could be viewed as the reflection back on the
brane of the gravitational field of $M$ by the negative AdS5 bulk cosmological
constant. In the limit $r < l$, it
can be shown that\cite{dadhich,dadhich1}
\be
Q = -\frac{Ml}{M_4^2}
\label{masscharge}
\ee
The bulk tidal charge thus strengthens the gravitational field of
the black hole. It has been further argued\cite{dadhich2} that since the 
back reaction of the bulk onto the brane strengthens gravity on the brane, 
the formation of a black hole as result of gravitational collapse is favored
as against a naked singularity.
The metric with negative tidal charge (\ref{tidalmetric})
has a spacelike singularity
and one horizon given by
\be
r_h = \frac{M}{M_4^2}\Biggl(1+ \sqrt{1 - \frac{QM_4^4}{M^2}}\Biggr)
\ee
which is larger than the Schwarzschild horizon. So the bulk effects are
seen to increase the entropy and decrease the temperature of the black
hole. 

A more general class of spherically symmetric and static solutions
to the field equations with a $5$-dimensional cosmological constant can be
derived by considering a general line element of the type
\be
ds^2 = -A(r)dt^2 + B(r)dr^2 + r^2(d\Omega^2)
\label{general}
\ee
and relaxing the condition $A(r)=B^{-1}(r)$ used while obtaining the
Schwarzschild or the Reissner-Nordstrom metrics. Casadio et al\cite{casadio}
obtained two types of solutions by fixing either $A(r)$ or $B(r)$, and
then demanding the correct $1/r$ asymptotic behaviour for the other
in terms of the post Newtonian (PPN) parametrization.
In the first case, the choice $A(r)= 1-2M/(M_4^2r)$ leads to the metric
\be
ds_4^2= -(1-\frac{2M}{M_4^2r})dt^2 + \frac{1-\frac{3M}{2M_4^2r}}
{(1-\frac{2M}{M_4^2r})
\left(1-\frac{M(4\beta-1)}{2M_4^2r}\right)} + r^2(d\Omega^2)
\label{casad1}
\ee
in terms of the PPN parameter $\beta$ which impacts the deflection and
time delay of light\cite{will}. Note that the above metric was also
derived as a possible geometry outside a star on the brane\cite{germani1}.
The solution (\ref{casad1}) is of the temporal Schwarzschild form having
a horizon $r_h = 2M/M_4^2$. The corresponding Hawking temperature is
given by\cite{casadio}
\be
T_{BH} = \frac{\sqrt{1-6(\beta-1)}}{8\pi M}
\ee
Thus, in comparison with Schwarzschild black holes, the black hole 
(\ref{casad1}) will be either hotter or colder depending upon the
sign of $(\b - 1)$.

Alternately, the choice for $B(r)$ of the form $B^{-1}(r) = 1- 2\gamma M/(M_4^2r)$,
in terms of the PPN parameter $\gamma$, yeilds the line element
\be
dS_4^2 = \frac{1}{\gamma^2}\Biggl(\gamma - 1 + \sqrt{1 - \frac{2M}{M_4^2r}}
\Biggr)^2
dt^2 + \frac{dr^2}{1-\frac{2M}{M_4^2r}} + r^2(d\Omega^2)
\label{casad2}
\ee
This form of the metric represents a nonsingular wormhole, and has been 
discussed earlier in the literature in the context of $4$-dimensional general 
relativity\cite{dadhich3}. Wormhole solutions in the braneworld context
have been discussed by Bronnikov et al\cite{bronnikov}.
Furthermore, a class of static, spherically symmetric and
non-singular braneworld solutions with horizon have 
been obtained\cite{dadhich4} by
relaxing the vanishing scalar curvature condition (\ref{fieldsoln}) used 
to obtain the solutions (\ref{tidalmetric}),(\ref{casad1}) and (\ref{casad2}).
Stationary solutions representing charged rotating black holes have also
been found recently\cite{aliev}.
The arbitrariness of the projected bulk Weyl term $E_{\mu\nu}$ and its
geometric origin is at the
root of the variety of braneworld black hole and wormhole solutions since 
both the
functions $A(r)$ and $B(r)$ in Eq.(\ref{general}) have to be determined
by it\cite{visser}. A specific configuration for the Weyl term with a negative
equation of state has been considered and the resultant geometry 
with a singular horizon has been worked
out to provide one more example of a possible braneworld black hole
solution\cite{gregory2}.

The black hole solution (\ref{tidalmetric}) exhibits a $1/r^2$ correction
to the Newtonian potential on the brane  in contrast to the weak field
correction of $1/r^3$ as in the solution (\ref{gartanmetric}). The
solution with tidal charge (\ref{tidalmetric}) is reflective of the
short distance or strong gravity limit where the $1/r^2$ correction
to the gravitational potential may even dominate over the $1/r$. This
corresponds to the fact that at short distances, braneworld gravity
is truly $5$-dimensional. For short distances $r \ll l$ it is 
natural to consider the $5$-dimensional Schwarzschild solution as a braneworld
black hole candidate given by\cite{myers}
\be
ds_5^2 = -\left(1-\frac{r_{BH}^2}{r^2}\right)dt^2+
\left(1-\frac{r_{BH}^2}{r^2}\right)^{-1}dr^2\\+r^2\left(d\Omega_3^2\right)
\label{hdbh}
\ee
where the horizon size $r_0$ is so small ($r_0 \ll l$) so that the
black hole effectively ``sees'' all the spatial dimensions on the same
footing. A generalysation to higher dimensions\cite{argyres,nakao} of the 
hoop conjecture leads to the above form of the metric as a static solution to
collapsing matter on the brane. Near the event horizon, the black hole
would have no way of distinguishing between the bulk dimension and the
braneworld ones. Numerical simulations for scales sufficiently small
compared to the AdS scale $l$ seem to also support the existence
of static solutions satisfying the AdS5 boundary conditions\cite{kudoh}.   

The induced $4$-dimensional metric on the brane near the event horizon of the
$5$-dimensional black hole (\ref{hdbh}) is obtained by integrating out 
the extra 
dimension to be
\be
dS_{4}^2=-\left(1-\frac{r_{BH}^2}{r^2}\right)dt^2+\left(1-\frac{r_{BH}^2}{r^2}\right)^{-1}dr^2\\+r^2\left(d\Omega^2\right)
\label{smallmetric}
\ee
This $4$-dimensional metric is different from the standard $4$-dimensional 
Schwarzschild solution
as it reflects the $5$-dimensional character of the strong gravitational 
field near
the black hole horizon in the form of the $1/r^2$ gravitational potential.
It is however expected that far from the event horizon the 
metric (\ref{smallmetric}) would approach the standard $4$-dimensional 
Schwarzschild
form, as was shown explicitly in $2+1$ dimensional braneworld 
framework\cite{emparan}. The properties of small black holes with the
geometry given by Eq.(\ref{smallmetric}) have been studied extensively by
Argyres et al\cite{argyres}. In general these black holes have 
lesser temperature and a longer lifetime 
compared to the standard $4$-dimensional 
Schwarzschild
black holes. Also since the $5$-dimensional Planck mass could be 
much lower compared
to the $4$-dimensional Planck mass ($M_5 \ll M_4$), these black holes could be
produced in particle accelerators\cite{giddings,dimopoulos} and cosmic ray 
showers\cite{feng}. They could thus
provide one avenue  of testing higher dimensional or braneworld
physics. Of course, the consequences of a population of primordial
black holes of the type (\ref{smallmetric}) are potentially rich during
many different cosmological eras, and these will be described in details in
section 4.

The Myers-Perry black hole\cite{myers} in $5$ dimensions (\ref{hdbh}) has been
also used in the context of braneworld models to investigate various effects
of its interaction with the brane and its radiation onto the bulk.
Frolov and Stojkovic\cite{frolov1} have shown that a small black hole 
attached to the brane may leave the brane as the result of a recoil
due to emission of quanta into the bulk. Such an effect leads to energy
loss in the brane. This opens up the possibility of observing energy
non-conservation in particle colliders which may be able to produce
these black holes. Radiation by rotating $5$-dimensional black holes have also
been studied and certain conditions have been found when such objects
could be stationary\cite{frolov2}. The interaction of $5$-dimensional black
holes with the brane described as a domain wall has interesting 
phenomenological features.
In particular, the induced geometry on the brane due to a moving
bulk black hole has been derived, and an apparent violation of the energy
condition observed on the brane\cite{frolov3}. Specific features of the
interaction of rotating black holes with the brane have been 
studied\cite{frolov4}. Energy flux through
the horizons of various configurations of the black hole--domain wall
system have also been investigated\cite{stojkovic1}. 

Before concluding this section, it needs to be emphasized that although
several spherically symmetric and static brane black hole solutions with 
contributions from the bulk gravity effects have been 
found\cite{dadhich,casadio,germani1,dadhich4,visser,gregory2}, and further
possibilities have been elucidated as belonging to a more general class
of black holes\cite{bronnikov}, none of
these are obtained as exact solutions of the full $5$-dimensional 
bulk field equations.
The analyses of gravitational collapse on the brane have typically
yeilded non-static solutions\cite{germani}.
Numerical simulations\cite{kudoh,numerical} have been  inconclusive in this 
aspect. Investigations on stellar metrics\cite{visser,wiseman} on the 
braneworld have
been performed in order to obtain further insight into the full $5$-dimensional
spacetime geometry. Understanding the bulk properties have been attempted
by extending some particular braneworld black hole
solutions to the bulk\cite{kanti2}.
The problem of finding the bulk metric which would represent
a static and spherically symmetric vacuum solution with horizon on the 
brane remains an open one till date.

\section{Braneworld cosmology}

The cosmology of the RS-II model entails a modified high energy phase
in the early radiation dominated era of the universe during which the
right hand side of the Einstein equation contains a term that is quadratic
in the brane energy momentum tensor\cite{maartens}. Other
modifications include the so-called
``dark-energy'' term which is given by the projection of the bulk Weyl
tensor. Transition to the standard radiation dominated era takes place
when $t >> t_c \equiv l/2$. Such a modified high energy evolution has
rich consequences for the physics of the early universe\cite{brax}.
In particular,
the inflationary scenario is altered, allowing the possibility of steep
inflaton potentials to accomplish the desired features.
Constraints on the duration of the brane dominated high energy phase
are enforced by the necessity of conforming to the standard cosmological
observational features such as nucleosynthesis and density perturbations.

Let us now review briefly some of the essential features of the
RS-II braneworld cosmology. The effective $4$-dimensional Einstein tensor
on the brane is given by\cite{maartens}
\be
G_{\mu\nu} = {8\pi\over M_4^2}\tau_{\mu\nu} + \k^4\Pi_{\mu\nu} - E_{\mu\nu}
\ee
where $\tau_{\mu\nu}$ is the  brane energy-momentum tensor; $\Pi_{\mu\nu}$ is
quadratic in
the brane EM tensor; and $E_{\mu\nu}$ is the projection of the 5-dimensional 
Weyl tensor.
The $4$-dimensional Planck's mass $M_4$ is related to the gravitational
coupling constant $\k$ and the AdS length $l$ by
$\frac{8\pi}{M_4^2} = \frac{\k^2}{l}$.

For the Friedmann-Robertson-Walker metric on the brane, the Friedmann
equation is given by
\be
H^2 = {8\pi\over 3M_4^2}\Biggl(\rho + {\rho^2\over 2\l} + \rho_{KK}\Biggr)
+ {\L_4\over 3} -{k\over a^2}
\ee
with $H$ being the Hubble constant, $\rho$ the energy density, and $k=-1,0,1$
representing open, flat and closed branes, respectively. $\rho_{KK}$ is
the effective energy density coming from the bulk Weyl tensor,
$\l \equiv 3M_5^6/4\pi M_4^2$ is the
brane tension, and $\L_4$ the effective $4$-dimensional cosmological 
constant on the
brane. The AdS curvature radius $l$ is given by the bulk cosmological
constant $\L_5$ and $5$-dimensional Planck mass $M_5$ as 
$\L_5 = -(3M_5^3)/(4\pi l^2)$.
The induced $4$-dimensional cosmological constant $\L_4$ is given by
\be
\L_4= 3\biggl({M_5^6\over M_4^4} - {1\over l^2}\biggr)
\label{cosmconst}
\ee
Setting $\L_4 =0$, one obtains a relation between the brane tension and
AdS radius given by
\be
\l^{-1/4} = \Biggl({4\pi\over 3}\Biggr)^{1/4}\Biggl({l\over l_4}\Biggr)^{1/2}
l_4
\label{tenserad}
\ee
Nucleosynthesis and CMBR observations constrain the ``dark energy'' 
term $\rho_{KK}$ to be
negligible compared to the radiation density $\rho$\cite{binetruy}.
For much of cosmological evolution one can neglect $\rho_{KK}$, as we will do
in the following analysis.

Assuming a radiation dominated equation of state, the $(k=0)$ solutions for
the Friedmann equation are given by
\be
\rho_R = {3M_4^2\over 32\pi t(t+t_c)}
\ee
for the energy density, and
\be
a=a_0\Biggl[{t(t+t_c)\over t_0(t_0+t_c)}\Biggr]^{1/4}
\ee
for the scale factor $a$ during the radiation dominated era, and where
$t_c \equiv l/2$ effectively demarcates the brane dominated ``high energy''
era from the standard radiation dominated era.
For times earlier that $t_c$, i.e., $t \le t_c$ (or $\rho \ge \l$), one has
the non-standard high energy regime during which the radiation density
and the scale factor evolve as
\be
\rho_R = {3M_4^2\over 32\pi t_ct}
\label{rhobrane}
\ee
and
\be
a = a_0\biggl({t\over t_0}\biggr)^{1/4}
\label{abrane}
\ee
respectively. As a consequence, the time-temperature relation also gets
modified during the brane dominated high energy era, i.e.,
$T \propto t^{-1/4}$. 

On the other hand, if the high energy braneworld regime
is matter dominated from a time $t_m$, the scale factor grows subsequently
like
\be
a= a_m\biggl({t\over t_m}\biggr)^{1/3}
\ee
But, for times much later than $t_c$, i.e.,
$t >> t_c$ (or $\rho << \l$), one should recover back the standard 
radiation dominated
cosmological evolution given by
\be
\rho_R = {3M_4^2\over 32\pi t^2}
\ee
and
\be
a = a_0\biggl({t\over (t_0t_c)^{1/2}}\biggr)^{1/2}
\ee
The observational success of standard big-bang nucleosynthesis constrains
that the high energy era be over by the epoch of the synthesis of light
elements. However, this requirement is satisfied for $l < 10^{43}l_4$,
which is a much weaker bound than that obtained from experiments
probing the modifications to Newtonian gravity\cite{long} in the braneworld
scenario.

Modified expansion in the high energy era has interesting implications
for inflation\cite{lidsey}. For inflation driven by a scalar field $\phi$ 
with potential $V$ on
the brane, the condition for accelerated expansion of the scale factor
($\ddot{a} > 0$) is satisfied when the equation of state parameter $w$ is
\be
w < -\frac{1}{3}\biggl(\frac{1+2\rho/\lambda}{1+\rho/\lambda}\biggr)
\ee
In the standard slow role approximation the Hubble rate and the scalar field
evolve as
\ben
H^2 \approx \frac{\kappa^2}{3}V\biggl(1 + \frac{V}{2\lambda}\biggr)
\dot{\phi} \approx \frac{V'}{3H}
\een
The braneworld correction term $V/2\lambda$ enhances the Hubble rate
compared to standard cosmology. This is turn increases friction in the
scalar field equation. Thus slow roll inflation is favored even for steep 
potentials\cite{cline}. The added advantage of such a scenario
is that the inflaton field can play the role of quintessence\cite{copeland} 
leading to a late time acceleration of the universe, as well. 
Further modifications to the high energy expansion in the very early
stages can be brought about by the inclusion of the Gauss-Bonnet term
in the $5$-dimensional action\cite{charmousis}. During the very early era the 
Gauss-Bonnet term drives the Hubble rate as $H^2 \propto \rho^{2/3}$
(Gauss-Bonnett regime), which subsequently changes to $H^2 \propto \rho^2$
(Randall-Sundrum regime), and finally for $t > t_c$ the $H^2 \propto \rho$
(standard regime)  evolution is recovered.  
Braneworld inflation could be  accomplished by the Gauss-Bonnet term for
very steep potentials such 
as the exponential potential\cite{tsujikawa}.

\section{Cosmological evolution of black holes}

This section will focus on the consequences of a population
of primordial black holes on  cosmology in the braneworld scenario.
The modified features of cosmology during the high energy braneworld
era have been highlighted in section 2. The  black holes present
in the early universe
affect the dynamics of the radiation dominated expansion through
Hawking emission and accretion of the surrounding 
radiation\cite{majumdar2,clancy}. These
two competing processes lead to a net energy flow for a single black hole,
the direction of which determines its longevity. The temperature and the 
rate of Hawking
radiation for braneworld black holes are themselves different from those of 
standard $4$-dimensional
Schwarzschild black holes. The evolution of the individual black
holes  are impacted by the accretion of radiation from
the surroundings, which is more effective in the braneworld scenario compared
to the standard cosmology. Since the 
rate of accretion is governed by the
rate of background expansion which is much slower in the high energy regime,
black holes that are produced earlier undergo larger growth. 

The actual rate of accretion and evaporation  of course depends upon the
particular geometry of the braneworld black hole. The black holes of
interest for cosmological evolution are produced very early in the
universe either due to the collapse of overdense regions resulting from
inflation generated density perturbations, or due to the collision of
heavy particles in the primordial plasma. Most of such black holes are
expected to be formed with a size small enough ($r \le l$) for the induced
$4$-dimensional Myers-Perry metric (\ref{smallmetric}) to be a good 
approximation to 
their geometry on the brane. Further, as we are interested in the processes 
of Hawking
evaporation and the accretion of radiation, both of which are characterized
by the near horizon short distance properties of the metric, it may also
be pragmatic to consider the $5$-dimensional form of gravity reflected
in the near horizon strong field region by the geometry in 
Eq.(\ref{smallmetric}). It is worth noting that the collapse of the
`tidal charge' in vacuum could also give rise to the same geometry 
for primordial black holes\cite{dadhich} ($M=0$ in Eq.(\ref{tidalmetric})).
For these reasons only this
particular form (\ref{smallmetric}) of braneworld black hole geometry 
has been considered for the analysis of cosmological 
evolution\cite{guedens,majumdar2,clancy,liddle}.

We first consider the evolution of a single primordial
black hole which is formed with a sub-horizon mass\cite{guedens} 
in the high energy
radiation dominated era. Since we are considering the $4$-dimensional 
projection
of the $5$-dimensional Schwarzschild metric, i.e.,
\be
dS_{4}^2=-\left(1-\frac{r_{BH}^2}{r^2}\right)dt^2+\left(1-\frac{r_{BH}^2}{r^2}\right)^{-1}dr^2\\+r^2\left(d\Omega^2\right)
\label{metric1}
\ee
the horizon radius of such a black hole is proportional to the square
root of its mass.
The mass-radius relationship given by
\be
r_{BH} = \Biggl({8\over 3\pi}\Biggr)^{1/2}\Biggl({l\over l_4}\Biggr)^{1/2}
\Biggl({M\over M_4}\Biggr)^{1/2}l_4
\label{massradius}
\ee
which is different from the ordinary $4$-dimensional Schwarzschild radius. 
Various properties of such black
holes have been elaborated\cite{argyres}, and the process of Hawking
evaporation into the bulk and also on the brane have been extensively
studied\cite{kanti}.
The Hawking evaporation rate, as for the case of standard black holes, is
proportional to the surface
area times the fourth power of temperature. The Hawking temperature
is given by
\be
T_{BH} = {1\over 2\pi r_{BH}}
\label{tempbbh}
\ee
Therefore, as a consequence of the mass-radius relationship (\ref{massradius}),
such black holes are colder and long-lived compared to $4$-dimensional
Schwarzschild black holes.

A black hole formed in the early radiation dominated era of the universe
accretes the surrounding radiation.
In standard cosmology the effectiveness of accretion in the growth of
black hole mass is restricted because of the fact that the mass 
of the individual black holes could grow at nearly the same rate as that
of the cosmological
Hubble mass $M_H$, 
i.e., $M_H \sim M \sim t$. Black holes produced in the radiation dominated
era cannot be formed with a size much smaller than the Hubble radius,
since otherwise pressure forces could hinder the collapse process.
However, in the radiation dominated high energy phase of the braneworld
scenario, the Hubble mass grows as $M_H \sim t^2$, whereas
the growth of  black hole mass due to accretion is given by $M \sim t^B$
(with $B < 2/pi$)\cite{majumdar2,clancy}. Hence, sufficient energy is available
within the Hubble volume for a black hole to accrete in the high energy
braneworld regime. The exact efficiency of accretion though depends upon
complex physical processes involving the mean free paths of the particles
comprising the radiation background and the thermal properties of the
radiation in the non-trivial geometry near the event horizon\cite{clancy}.
Any peculiar velocity of the black hole with respect to the cosmic frame
further impacts the rate of accretion. In the absence of a universally
accepted approach of determining the precise accretion rate, it is
usually taken to be as equal to the product of the 
surface area of the black hole, the energy density of the background
radiation, and an efficiency factor ranging between $0$ and 
$1$\cite{clancy}. 

Taking into account these
effects of accretion and evaporation together, the rate of change of
mass $\dot{M}$ of a braneworld black
hole is given by
\be
\label{bbhrate}
\dot{M} = 4\pi r_{BH}^2\biggl( - g_{brane}\sigma T_{BH}^4 + f\rho_R\biggr)
\ee
where
$g_{brane}$ is effective number of particles that can be emitted by
the black hole (we assume that the black holes can emit massless particles
only and take $g_{brane} = 7.25$\cite{guedens}),
$f$ is the accretion efficiency ($0 \le f \le 1$)\cite{clancy}, and
 $\sigma$ is the Stefan-Boltzmann constant.
The black hole also evaporates into
the bulk, with a rate  proportional to $4\pi r_{BH}^2g_{bulk}T_{BH}^5$.
However, this term is subdominant even for very small black
holes\cite{guedens}, and has negligible effect on their lifetimes.

Substituting the expressions for the black hole radius (Eq.(\ref{massradius})),
the temperature-radius relation (\ref{tempbbh}), and the energy density of
radiation (\ref{rhobrane}), the black hole rate equation (\ref{bbhrate}) in
the radiation dominated high energy braneworld era can be
written as\cite{majumdar2}
\be
\dot{M} = -{AM_4^2 \over Mt_c} + {BM \over t}
\label{bbhrate2}
\ee
where $A$ and $B$
are dimensionless numbers given by
\ben
\label{AandB}
A &\simeq & {3\over (16)^3\pi} \\
B &\simeq & {2f\over \pi}
\een
The exact solution for the black hole rate equation is given by\cite{majumdar2}
\be
\label{exactsoln}
M(t) = \Biggl[\Biggl(M_0^2 - {2AM_4^2\over 2B -1}{t_0\over t_c}\Biggr)
\Biggl({t\over t_0}\Biggr)^{2B} + {2AM_4^2\over 2B -1}{t\over t_c}\Biggr]^{1/2}
\ee
with $M_0$ being the formation mass of the black hole at time $t=t_0$.
If the black hole is formed out of the collapse of horizon or sub-horizon sized
density perturbations, the formation time and mass are related by\cite{guedens}
\be
\label{initmasstime}
{t_0\over t_4} \simeq {1\over 4}\Biggl({M_0\over M_4}\Biggr)^{1/2}\Biggl({l\over l_4}\Biggr)^{1/2}
\ee
It has been argued\cite{majumdar2} that a black hole so formed continues
to grow in size by the accretion of radiation during the high energy
radiation dominated era, with its mass increasing as
\be
\label{massgrowth}
{M(t) \over M_0} \simeq \Biggl({t\over t_0}\Biggr)^B
\ee
This result is however sensitive to the accretion efficiency, since a more 
careful analysis\cite{clancy}  shows that for $f < \pi/4$, the growth
due to accretion could come to a halt during the high energy regime itself.

Within the context of standard cosmology, the evolution of a population 
of primordial
black holes exchanging energy with the surrounding radiation by accretion
and evaporation has been studied by several authors\cite{majumdar1,custodio}. 
The analysis of this problem is  simplified by assuming 
that all the black holes are
formed with an average initial mass $M_0$ at
a time $t_0$ when the fraction of the total energy density in black holes is
$\b_{BH}$, and the number density of black holes is $n_{BH}(t_0)$. With
these assumptions,
the coupled cosmological equations for the radiation density
$\rho_R(t)$, the matter density in the black holes $M(t)n_{BH}(t)$, and the
scale factor $a(t)$ are integrated to give the complete cosmological
evolution. Note however,  that the black
holes are produced with an initial
mass spectrum in any realistic scenario of black
hole formation in the standard cosmology\cite{starobinsky}, and it 
is expected that the same would be
the case in the braneworld scenario as well. 
The effect of a
mass distribution can be incorporated into the cosmological
evolution in the standard scenario
by introducing an additional differential equation for the distribution
function and specifying further initial conditions related to 
it\cite{custodio}. However, since not much is known about the formation
processes of black holes in braneworld cosmology, the study of only a few
basic features of their cosmological evolution under simplifying
assumptions has been undertaken in
the literature till date.

The number density of black holes $n_{BH}(t)$ scales as $a(t)^{-3}$, and
thus for a radiation dominated evolution on the brane, one gets
\be
(n_{BH}(t)/n_{BH}(t_0)) = (t_0/t)^{3/4}
\ee
 since $a(t) \propto t^{1/4}$.
The net energy in black holes grows since accretion dominates over
evaporation. The condition for
the universe to remain radiation dominated (i.e., $\rho_{BH}(t) < \rho_R(t)$)
at any instant $t$ can be derived
to be\cite{majumdar2}
\be
\label{raddomcond}
\b_{BH} < {(t_0/t)^{B+1/4} \over 1 + (t_0/t)^{B+1/4}}
\ee
If the value of $\b_{BH}$ exceeds the above bound, there ensues an era of
matter (black hole) domination in the high energy braneworld phase. Such
a phase of matter domination should definitely be over by the time
of nucleosynthesis for the cosmology to be viable.

Let us first describe the situation when the cosmology stays radiation
dominated up to the time when brane effects are important, i.e., $t\leq t_c$.
From Eq.(\ref{raddomcond}), this requires
\be
\label{raddomcond2}
{\b_{BH} \over 1 -\b_{BH}} < \Biggl({t_0 \over t_c}\Biggr)^{B + 1/4}
\ee
Further, the black holes should
remain small enough, i.e.,
$(M/M_4) < (3\pi/4)(t/t_4)$
for the $5$-dimensional evaporation law to be valid\cite{guedens}. These 
criteria
can be used to put an upper bound to the average initial mass\cite{majumdar2}
\be
\label{masslim}
{M_0\over M_4} < \Biggl({3\pi \over 4(2\sqrt{2})^B}\Biggr)^{{2\over 2-B}}
{t_c \over t_4}
\ee

The growth of the black holes in the radiation dominated
era due
to accretion slows down with time, since the surrounding radiation density
gets diluted.
The rate of evaporation is also insignificant for a wide range of $M_0$ at
this stage since the
black hole masses could
have grown by several orders of magnitude from their initial values.
There ensues
an era during which the black hole mass stays nearly constant over a period
of time, as is the case for standard cosmology\cite{majumdar1}.
The accretion
rate is smaller for the braneworld case since the surface area is
proportional to $M$ instead of $M^2$
for $4$-dimensional black holes. Moreover, the
 evaporation rate is ($\propto M^{-1}$) instead of $M^{-2}$. Hence, the
black hole mass will stay for while near a maximum value $M_{max}$ reached
at time $t_t$ before evaporation
starts dominating. The expression for the lifetime $t_{end}$ of a black hole in
this scenario is given by\cite{majumdar2}
\be
\label{lifetime}
{t_{end} \over t_4} \simeq {4\over A}(2\sqrt{2})^B\Biggl({M_0\over M_4}\Biggr)^{2-B}
{t_c\over t_4}\Biggl({t_t^2 \over t_c t_4}\Biggr)^B
\ee
It is important to note that the modified evaporation law also contributes
to the increased lifetime for braneworld black holes\cite{guedens}.
However, the effect of accretion is more significant, as can be
seen by comparing the lifetime of a $5$-dimensional black hole in the presence
of accretion to the the lifetime due to purely an altered
geometry\cite{majumdar2}.
Depending upon the values taken by the parameters $t_c$ and $l$, one
could obtain several interesting examples of primordial black holes
surviving till various cosmologically interesting eras\cite{majumdar2,clancy}.
One particular choice worth mentioning is that of a black hole formed
with an initial mass $M_0 = 10^8 M_4 \simeq 10^3{\rm g}$, that will survive
up to the present era if one chooses $(l/l_4) \simeq 10^{30}$.

Further interesting phenomena occur in the interaction of two or more 
neighbouring black holes mediated by the surrounding radiation
in the radiation dominated high energy era. These
black holes exchange energy via the processes of evaporation and accretion
with the radiation bath, and through it, with each other. The evolution
equation for a black hole gets modified due to the presence of one
or more neighbouring black hole(s) with the addition of an extra source (sink)
term due to the evaporating (accreting) neighbour(s) over the average
radiation background. The black hole equation (\ref{bbhrate2}) now becomes
\be
\dot{m}_i = {Bm_i\over \tilde{t}} -{1\over m_i} - g{m_i\dot{m}_j\over
\tilde{t}^{1/2}}
\label{evolve2}
\ee
where $\tilde{t}=Am_4^2t/t_c$, and
\be
g = {4A^{1/2}\over 3\pi}\biggl({l_4\over d_0}\biggr)^2\biggl({t_0\over
t_4}\biggr)^{1/2}\biggl({t_c\over t_4}\biggr)^{1/2}
\label{coupling}
\ee
can be dubbed as the `coupling' parameter between two black holes.
The
physical distance between
two such neighbours increases initially with the Hubble expansion.
Forming out of horizon collapse, the initial mass ratio of two such black
holes is on average proportional to the square of the ratio of their
formation times (from Eq.(\ref{initmasstime})). By sudying the effect of
the interaction
term it is possible to show that an initial mass difference between two
such black
holes can never decrease during the radiation dominated era\cite{majumdar3}.

The energy exchange between the black holes and the surrounding radiation
always causes mass disequilibration between neighbours.
Such mass differences  facilitate the formation of binaries
during the standard low energy phase via three-body gravitational
interactions. A formation mechanism for primordial black hole binaries
in standard cosmology has been investigated\cite{nakamura}
which shows that if equal mass primordial black holes are
inhomogeneously distributed
in space, three-body gravitational interactions could lead to the formation
of binaries.
The scheme of binary formation in the braneworld scenario\cite{majumdar3}
could become more effective
when combined with the scheme  based on spatial
inhomogeneities discussed in the context of standard cosmology\cite{nakamura}.
Coalescing braneworld black hole
binaries may emit gravitational waves amenable for detection by the next
generation detectors\cite{inoue}. 

Let us now describe the case of matter (black hole)
domination  in the high energy
phase. The onset of such a matter dominated era is
derived to be\cite{majumdar2}
\be
\label{mattdomtime}
{t_{heq}\over t_0} = \Biggl({1-\b_{BH} \over \b_{BH}}\Biggr)^{{4\over 4B+1}} \equiv \g_B
\ee
The mass of a black hole at $t_{heq}$ is given by $M(t_{heq}/M_0) = \g_B^B$.
For $t > t_{heq}$ the Hubble expansion is essentially driven by the black
holes ($p=0$) which
dominate over radiation. Since the number density of black holes scales
as matter ($n_{BH}(t) \propto a^{-3}$), for $t<t_c$ one has
$H \propto \rho_{BH}$,
and thus the scale factor grows as
\be
a(t) \sim t^{1/3}
\ee
During this era, the
radiation density $\rho_R$ is governed by the equation
\be
{d \over dt}\biggl(\rho_R(t)a^4(t)\biggr) = - \dot{M}(t)n_{BH}(t)a(t)
\ee
where the contribution from accreting black holes is
comparable to the normal redshiting term ($\rho_R \sim a^{-4}$)
because at this stage the black holes dominate the total energy density.
Some further analysis leads to the following expression for
the radiation density\cite{majumdar2}
\be
\rho_R(t) \approx \g_B^{-1}\rho(t_0)\Biggl({t_{heq}\over t}\Biggr) - {\b_{BH} B\over
B+1/3}
\g_B^{1/4}\rho(t_0)\Biggl({t_0\over t}\Biggr)^{1-B}
\ee
The black hole mass grows as
\be
M(t) = M_0\g_B^B{\rm exp}\Biggl[3B + {C\over B}\g_B^{B+1/4} - 3B\Biggl({t_{heq}\over t}\Biggr)^{1/3} - \g_B^{1/4}{C\over B}\Biggl({t\over t_0}\Biggr)^B\Biggr]
\ee
where $C = \frac{\b_{BH} B}{B+1/3}$. As in the case of standard cosmology, 
the accretion regime 
lasts for a
brief duration in the matter dominated
phase, beyond which the black hole evaporation starts to play a
significant role.

The universe gets reheated as $\rho_R(t)$ increases with time. The stage of
black hole domination lasts up to a time $t_r$
($\rho_R(t_r) = n_{BH}(t_r)M(t_r)$).
Subsequently, radiation domination takes over once again.
One can derive\cite{majumdar2}
\be
{t_c \over t_r} \approx {3\over 2\d_B\g_B} + {A\over 4\g_B^{2B}}\Biggl({M_4\over M_0}\Biggr)^2
\ee
with
\be
\d_B \equiv \Biggl({B+1/3 \over 1-\b_{BH}}\Biggr)^{{3\over 3\b_{BH} +1}}
\ee
A stringent restriction is imposed by demanding that the standard low energy
cosmology (for $t > t_c$) should emerge as radiation
dominated. By requiring that the era of black hole
domination be over before $t_c$, i.e., $t_r < t_c$, one gets a lower
bound on $\b_{BH}$ from Eq.(35), i.e.,
\be
\b_{BH} \geq \Biggl[{4t_0\over 3t_c}\bigl(B + 1/3\bigr)^{{3\over 3B+1}}\Biggr]^{B+1/4}
\ee
The evaporation time of black holes in this scenario has also been
calculated\cite{majumdar2}.
The black hole
lifetime is given by
\be
\label{lifetime2}
{t_{end}\over t_4} \approx \Biggl({M_0\over M_4}\Biggr)^2{t_c\over t_4}\g_B^{2B}\ee
The effect of accretion is less significant in prolonging black hole
lifetimes in this case, as borne out by the following example.
For instance, taking the value of the AdS radius to be $l/l_4 \simeq 10^{20}$,
and $\b_{BH} = 10^{-3}$, black holes with $M_0 \simeq 10^{12}{\rm g}$ evaporate
during the present era. This is only to be expected, since early matter
domination results in larger Hubble expansion rate which further restricts
the availability of radiation for the black holes
to accrete.

It thus turns out that it is possible to have primordial black holes
formed in the high energy era of the braneworld scenario to survive
up to several cosmologically interesting eras. Accretion of radiation
during the radiation dominated era is primarily responsible for
the increased longevity of braneworld black holes, though their $5$-dimensional
geometry also contributes to a slower rate of Hawking evaporation.
It is worth noting here that primordial black holes in certain
models could also accrete the energy of a cosmological
scalar field\cite{bean}.
Such an effect could lead to further growth of these black holes beyond
the early radiation dominated era, thus pushing up their lifetimes
further. The implications for
a population of primordial black holes in cosmology are diversely
manifold. At any particular era, the surviving black holes would contribute
a portion to the the total energy density as dark matter in the universe.
If the black holes are produced with an initial mass spectrum, then one
would have evaporating black holes at different eras. Hawking radiation from
these evaporating black holes would on one hand produce all kinds
of particles including heavier ones which could lead to 
baryogenesis\cite{majumdar1},
and on the other, could contribute significantly to the background
photons, thus diluting the baryon to photon ratio. 

Observational
constraints impacting different cosmological eras could be used to impose
restrictions on the initial mass spectrum of braneworld black holes in
a manner similar to the primordial black holes in standard cosmology.
Clancy et al\cite{liddle} have shown how standard constraints are modified
in the case of braneworld cosmology. 
To simplify the treatment, one can assume that radiation domination 
persists up to $t_c$, and that the accretion of radiation is possible 
only up to $t_c$.
A black hole mass fraction $\alpha_M$ in terms of the radiation density can be
defined (related to the mass fraction $\beta_{BH}$) as
\ben
\alpha_{M_0}(t) = \frac{\rho_{BH}(t)}{\rho_T} \nonumber \\
\alpha_{M_0}(t_0) = \frac{\beta_{BH}}{1-\beta_{BH}}
\een
Similarly, a `final' mass fraction $\alpha(t_{\mathrm{evap}})$ is defined
at the end of the black hole life-cycle, since the black holes radiate
most of their energy towards the very end of their lifetimes.
Observational constraints
on $\alpha(t_{\mathrm{evap}})$ are considered at different cosmological 
epochs, given by
\be
\alpha (t) < L_{4D}(t)
\ee
for standard $4$-dimensional black holes, or 
\be
\alpha (t) < L_{5D}(t)
\ee
for braneworld black holes.
Then these constraints are evolved
backwards to constrain initial mass spectrum.
Accretion in the high-energy phase leads to $\alpha (t) \propto M_{BH}(t)a(t)$.
Therefore the constraints on the initial black hole mass fraction 
are given by
\be
L_{4D}^0 \equiv \alpha_i < [\frac{a_0}{a(t)}]_{4D} L_{4D}(t)
\ee
for standard cosmological evolution, and 
\be
L_{5D}^0 \equiv \alpha_i < [\frac{a_0}{a(t)}]_{5D} L_{5D}(t)
\ee
in the braneworld scenario.
Any astrophysical or cosmological process to be constrained at certain epoch 
is dominantly affected  by the PBHs with lifetimes
of that epoch. The constraints on primordial black holes in standard
cosmology thus get modified to\cite{liddle}
\be
\frac{L_{5D}^0}{L_{4D}^0} = \frac{L_{5D}(t_{\mathrm{evap}})}{L_{4D}(t_{\mathrm{evap}})} \biggl(\frac{l}{l_{\mathrm{min}}}\biggr)^{\frac{5-16B}{16-8B}}
\ee
where $l_{\mathrm{min}} \propto t_{\mathrm{evap}}^{1/3}$.
The departure from standard constraints is sensitive to the accretion 
efficiency, as is expected.

A more detailed study on how the initial mass spectrum of the black
holes is distorted due to braneworld accretion has been undertaken by
Sendouda et al\cite{sendouda1}. The diffuse photon background emitted
by the spectrum of black holes has been shown to be modified in accordance
with the mass spectrum of the black holes. These results have been compared
to the observed diffuse photon background to obtain bounds on the initial
black hole mass fraction, the scale of the extra dimension, and the accretion
efficiency by these authors. The observed number density of massive 
particles could also be used to obtain bounds on the initial mass fraction
of the black holes, as in the case of Schwarzschild primordial black
holes in standard cosmology. Further constraints on the scale of the extra
dimension have been derived\cite{sendouda2} by considering the recent 
observation of sub-Gev galactic antiprotons\cite{orito} as originating from
braneworld black holes present in our galaxy. A relevant issue 
is to investigate whether
primordial braneworld black holes could contribute to a significant
fraction of cold dark matter. If so, the direct searches of cold dark matter
compact objects through gravitational lensing might be
able to reveal their presence in galactic haloes. In the next section we will 
report on the specific features of the analysis of gravitational lensing for
several braneworld black hole metrics.

\section{Gravitational lensing by braneworld black holes}

The braneworld scenario implies the modification of Einstein's general
relativity at short distances or strong gravitational fields.
The bending of light due to the gravitational potential of a massive
object is one of the first predictions of the general theory of relativity.
Its application in the phenomenon of gravitational
lensing\cite{schneider} has potentially diverse possibilities.
Gravitational lensing in the weak field limit\cite{bernard} is till date
one of the most widely used tool in observational astrophysics and
cosmology. On the other hand, strong field gravitational lensing, though
limited in observational utility because of presently inadequate instruments,
remains our ultimate scope for exploring the physics of strong gravitational
fields. The general technique for analysing strong gravitational lensing
for spherically symmetric metrics has been developed by 
Bozza\cite{bozza1,bozza}
who has also formulated useful connections between observational
quantities like fluxes and resolutions and the metric parameters.
Strong gravitational lensing
is also endowed with richer phenomenological features like relativistic
images\cite{virbhadra1,virbhadra2} and retrolensing\cite{retro,eiroa2,bozza2}.
It would be thus worthwhile to investigate if the results of strong
gravitational lensing could be used for probing the modifications to general
relativity made in braneworld geometries. Such studies are also motivated
from the possibility of producing braneworld black holes in future
accelerators\cite{giddings,dimopoulos}.

Analysis of the trajectories of light and massive particles
in various braneworld and higher dimensional metrics have been undertaken
recently. Kar and Sinha\cite{kar} obtained the bending angle of light for
several brane and bulk geometries. Frolov et al\cite{frolov5}
found certain non-trivial features about the propagation of light in the
Myers-Perry\cite{myers} metric for a $5$-dimensional black hole
solution. This solution represents primordial black holes that could
be produced with size $r < l$ in the early high energy era of the
RS-II braneworld
scenario\cite{guedens}. It was shown that such black holes could grow
in size due to accretion of radiation, and consequently survive till
much later stages in the evolution of the universe\cite{majumdar2}.
With a suitable choice of parameters, some of these black holes could
also exist in the form of coalescing binaries in galactic haloes at
present times\cite{majumdar3}. The weak field limit of gravitational lensing
was studied for the Myers-Perry metric and certain notable differences
from the standard Schwarzschild lensing were found\cite{majumdar4}.
Thereafter, Eiroa\cite{eiroa} analysed the strong field lensing and
retrolensing effects for the Myers-Perry black hole. Strong field
gravitational lensing in a couple of other braneworld metrics has been
discussed by Whisker\cite{whisker} and some lensing observables have been
computed using parameters for the galactic centre black hole.

For a spherically symmetric metric 
\begin{equation}
ds^2 = - A(r)dt^2  +B(r)dx^2  +C(r)\left(d\Omega^2\right)
\label{sphersym}
\end{equation}
where the asymptotic forms of the functions $A(r)$ and $B(r)$ have the
standard $1/r$ form, and $C(r) \to r^2$ asymptotically, the general 
formalism of strong field
gravitational lensing has been worked out by Bozza\cite{bozza}. It is
required that the equation
\begin{equation}
\frac{C'(r)}{C(r)}=\frac{A'(r)}{A(r)}
\label{photsphere}
\end{equation}
admits at least one positive solution the largest of which is defined
to be the photon sphere $r_{\mathrm ph}$. 

\begin{figure}[pb]
%\begin{center}
\centerline{\psfig{file=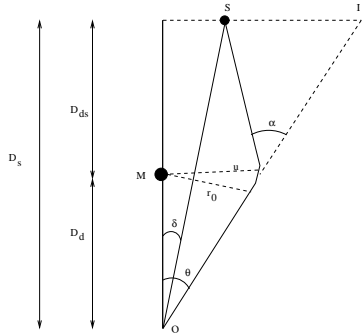,width=4.7cm}}
\vspace*{8pt}
\caption{Gravitationl lensing for point like mass object $M$. A light ray
from the
source $S$ passes the lens with an impact parameter $u$, and  is deflected by 
an angle $\alpha$.
The observer sees an image $I$ of the source at the angular position $\theta$.
\label{f1}}
%\end{center}
\end{figure}

A photon emanating from a 
distant source and having an impact parameter $u$ will approach near
the black hole at a minimum distance $r_0$ before emerging in a different
direction (see Figure~\ref{f1}). The closest approach distance is given in 
terms of the impact
parameter by
\begin{equation}
u=\sqrt{\frac{C_0}{A_0}}
\label{impact}
\end{equation}
where the functions $C$ and $A$ are evaluated at $r_0$. The deflection
angle of the photon in terms of the distance of closest approach is
\begin{eqnarray}
&& \alpha(r_0)=I(r_0)-\pi \label{deflangle1} \\
&& I(r_0)=\int\limits_{r_0}^\infty \frac{2\sqrt{B}dr}{\sqrt{C}
\sqrt{\frac{C}{C_0}\frac{A_0}{A}-1}}
\label{deflint1}
\end{eqnarray}

The weak field limit is obtained by expanding the integrand in 
Eq.(\ref{deflint1}) to the first order in the gravitational potential.
This limit however is not a good approximation when there is 
a significant difference
between the impact parameter $u$ and the distance of closest approach
$r_0$, which occurs when $A(r_0)$ significantly differs from $1$, or
$C(r_0)$ from $r_0^2$. By decreasing the impact parameter, and consequently
the distance of closest approach, the deflection angle increases beyond
$2\pi$ at some stage resulting in one or more photonic loops around
the black hole before emergence. Further decrease of the impact parameter
to a minimum value $u_m$ corresponding to the distance of closest
approach $r_0 = r_{\mathrm ph}$ results
in the divergence of the deflection angle (integral in Eq.(\ref{deflint1})),
which means that the photon is captured by the black hole. Strong field
gravitational lensing is useful for studying the deflection of light
in a region starting from just beyond the photon sphere up to the distance
where the weak field approximation approaches validity.

In the general analysis of strong field gravitational lensing it becomes
necessary to extract out the divergent part of the deflection angle. In
order to do so two new variables $y$ and $z$ are defined as\cite{bozza}
\begin{eqnarray}
&& y=A(r) \\
&& z= \frac{y-y_0}{1-y_0}
\label{newvar}
\end{eqnarray}
where $y_0=A_0$. In terms of these variables, the integral (\ref{deflint1}) 
in the deflection angle is given by
\begin{eqnarray}
&& I(r_0)=\int\limits_0^1 R(z,r_0) f(z,r_0) dz \label{I z}\label{deflint2} \\
&& R(z,r_0)=\frac{2\sqrt{B y}}{C A'}\left( 1-y_0 \right)
\sqrt{C_0} \label{R} \\
&& f(z,r_0)=\frac{1}{\sqrt{y_0- \left[ \left(1-y_0 \right) z+ y_0
\right]\frac{C_0}{C}}}
\end{eqnarray}
where all functions without the subscript $0$ are evaluated at
$r=A^{-1} \left[\left(1-y_0 \right) z+ y_0 \right]$. 
The function $R(z,r_0)$ is regular for all values of $z$ and $r_0$, while
$f(z,r_0)$ diverges for $z \to 0$.

The deflection angle can be written as a
function of $\theta = u/D_{d}$, where $\theta$ is the angular separation
of the image from the lens, and $D_{d}$ is the distance between the lense
and the observer (See Figure~\ref{f1}). 
In order to do so, an integral
\begin{equation}
b_R = \int\limits_0^1 g(z,r_m) dz
\label{deflint3}
\end{equation}
is defined, where
\begin{equation}
g(z,r_m)= R(z,r_m)f(z,r_m)-R(0,r_m)f_0(z,r_m)
\end{equation}
The expression for the deflection angle is given as a
function of $\theta = u/D_{d}$
as\cite{bozza}
\begin{eqnarray}
&& \alpha(\theta)=-\overline{a} \log \left( \frac{\theta
D_{d}}{u_m} -1
\right) +\overline{b} \label{deflangle2}\\%
&& \overline{a}= \frac{R(0,r_m)}{2\sqrt{\beta_m}} \label{abar}\\%
&& \overline{b}= -\pi+b_R+\overline{a}\log{\frac{2\beta_m}{y_m}}\label{bbar}
\end{eqnarray}
where $R(0,r_m)$ and $b_R$ are given be Eqs.(\ref{R}) and (\ref{deflint3}),
respectively, and $\beta_m$ is defined as
\begin{equation}
\beta_m=\frac{ C_m \left( 1- y_m \right)^2
\left(C''_m y_m-C_m A''(r_m) \right)}{2y_m^2 {C'_m}^2}
\label{betam}
\end{equation}

In strong lensing there may exist
$n$ relativistic images given by the number of times a light ray loops
around the black hole. The positions of these are obtained as solutions of 
the lense equation given by
\be
{\mathrm tan} \delta = {\mathrm tan} \theta - \frac{D_{ds}}{D_s} 
[{\mathrm tan} \theta + {\mathrm tan}(\alpha - \theta)]
\label{lenseq}
\ee
for specific positions of the source and the lense respective to the 
observer, and using the the value of the deflection angle from 
Eq.(\ref{deflangle2}). The relativistic images formed by light rays winding
around the black hole are highly demagnified compared to the weak field
images. When the source, the lense and the observer are highly aligned, it
is possible to obtain the most prominent of the relativistic 
images\cite{virbhadra1}. Hence, the analysis of strong lensing is usually
restricted to the case when both $\delta$ and $\theta$ are 
small\cite{bozza}, though the general case for arbitrary positions
can also be analysed\cite{bozza2}. With the above restriction on the values of
$\delta$ and $\theta$, a light ray will reach the observer after winding 
around the lense $n$ number of times only if the deflection angle $\alpha$ is
very close to a multiple of $2\pi$. Substituting 
$\alpha = 2n\pi + \Delta \alpha_n$ in Eq.(\ref{lenseq}), one gets
\be
\delta = \theta - \frac{D_{ds}}{D_s} \Delta \alpha_n
\label{lenseq2}
\ee

The position of the $n$-th relativistic image $\theta_n$ can hence be obtained 
as a solution of the lense equation (\ref{lenseq2}) as\cite{bozza}
\be
\theta_n = \frac{u_m}{D_d}(1 + e_n) + \frac{u_m e_n \biggl(\delta - 
\frac{u_m(1+e_n)}{D_d}\biggr)D_s}{\overline{a}D_{ds}D_d}
\label{nposit}
\ee
where $u_m$ is the minimum impact parameter, and $e_n$ is given by
\be
e_n = {\mathrm e}^{(\overline{b} -2n\pi)/\overline{a}}
\ee
The magnification $\mu_n$ of the $n$-th relativistic image is given 
by\cite{bozza}
\be
\mu_n = \frac{1}{(\delta/\theta)\partial \delta \partial \theta}
\vert_{\theta_n} \simeq \frac{u_m^2 e_n (1+e_n)D_s}{\overline{a}\delta 
D_{ds}D_d^2}
\label{magnif}
\ee
The above formula for magnification is valid under the approximation of
a point source. However, for an extended source the magnification at
the image positon can also be derived by integrating over the luminosity
profile of the source\cite{eiroa2}. 

It is useful to obtain the expressions for the various lensing observables 
in terms of the metric parameters.
For $n \to \infty$ an observable $\theta_{\infty}$
can be defined\cite{bozza} representing the asymptotic position approached
by a set of images. 
The minimum impact parameter can then be  obtained as
\begin{equation}
u_m=D_{d} \theta_{\infty}
\label{thetainfty}
\end{equation}
In the simplest situation where only the
outermost image $\theta_1$ is resolved as a single image, while
all the remaining ones are packed together at $\theta_\infty$, 
two lensing observables can be defined as\cite{bozza}
\begin{eqnarray}
{\cal S}=\theta_1-\theta_\infty 
\end{eqnarray}
representing the separation between the first
image and the others, and 
\begin{eqnarray}
{\cal R}=\frac{\mu_1}{\sum\limits_{n=2}^\infty \mu_n}
\end{eqnarray}
corresponding to the ratio between the flux of the first
image and the flux coming from all the other images. 

In terms of the deflection angle parameters
$\overline{a}$ and $\overline{b}$, these observables can be written
as\cite{bozza}
\begin{equation}
{\cal S}= \theta_\infty e^{\overline{b}/\overline{a}
- 2\pi/\overline{a}} 
\label{obs-s}
\end{equation}
\begin{equation}
{\cal R}=e^{2\pi/\overline{a}}
\label{obs-r}
\end{equation}
The above equations (\ref{obs-s}) and (\ref{obs-r}) can be inverted to express
$\overline{a}$ and $\overline{b}$ in terms of the image separation ${\cal S}$
and the flux ratio ${\cal R}$. Therefore the knowledge of these two observables
can be used to reconstruct the deflection angle given by Eq.(\ref{deflangle2}).
The aim of strong field gravitational lensing is to detect the 
relativistic images corresponding to specific lensing candidates and 
measure their separations and flux ratios. Once
this is accomplished, the observed data could be compared with the
theoretical coefficients obtained using various metrics. A precise set
of observational data for strong gravitational lensing, if obtained,  could 
therefore be able discriminate between different models of gravity.
In the braneworld scenario the computation of the above parameters has
been performed for two metrics\cite{whisker} taking the black hole
at the centre of our galaxy as a potential candidate. 

We will now consider examples of the various lensing quantities
defined above for some of the possible braneworld black hole geometries
discussed in section 2. It is instructive to compare the braneworld
lensing quantities with the standard Schwarzschild ones which for strong
gravitational lensing are given as follows\cite{bozza}. Choosing the
Schwarzschild radius $r_s = 2M/M_4^2$ as the unit of distance, the photon
sphere is given by
\be
r_{\mathrm ph} = \frac{3}{2}
\label{phot}
\ee
for Schwarzschild lensing. The corresponding minimum impact parameter is
\be
u_m = \frac{3\sqrt{3}}{2}
\label{minimp}
\ee
The coefficients $\overline{a}$ and $\overline{b}$ defined in Eqs.(\ref{abar})
and (\ref{bbar}) are given by
\ben
\overline{a} &=& 1 \\
\overline{b} &=& -\pi + 2 {\mathrm log}[6(2-\sqrt{3})] + {\mathrm log}[6]
\een
The deflection angle (\ref{deflangle2}) is obtained in terms of the 
parameters $\overline{a}$ and $\overline{b}$ to be\cite{bozza}
\be
\alpha(\theta) = -{\mathrm log}\biggl(\frac{2\theta D_d}{3\sqrt{3}}-1\biggr)
+ {\mathrm log}[216(7- 4\sqrt{3})] - \pi
\ee
In the weak field limit, the expression for the bending angle is given by
\be
\alpha_{\mathrm weak} = \frac{4M}{M_4^2 r_0}
\ee

For the analysis of gravitational lensing by braneworld metrics, let us
first consider the Garriga-Tanaka weak field solution\cite{garrtan}
given by Eq.({\ref{gartanmetric}) in isotropic coordinates. The metric in
terms of the standard coordinates can be written as
\be
ds_4^2=-\left(1-\frac{2M}{M_4^2r}+\frac{4Ml^2}{3M_4^2r^3}\right)dt^2+
\left[\frac{1+\frac{M}{M_4^2r}-\frac{Ml^2}{3M_4^2r^3}}{1+\frac{2M}{M_4^2r}
+\frac{2Ml^2}{3M_4^2r^3}}\right]^{-2}dr^2 +r^2(d\Omega^2)
\label{gartanmetric2}
\ee
The formal expression for the radius of the photon sphere $r_{\mathrm ph}$ 
can be 
obtained as a function of $M$ and $l$ using Eq.(\ref{photsphere}) to be
\ben
r_{\mathrm ph} = \frac{r_s}{2}
&&+\frac{3^{1/3}r_s^2}{2(3r_s^3-20l^2r_s+2\sqrt{10}
\sqrt{10l^4r_s^2-3l^2r_s^4})^{1/3}}\nonumber\\
&&+\frac{(3r_s^3-20l^2r_s+2\sqrt{10}\sqrt{10l^4r_s^2-3l^2r_s^4})^{1/3}}{2 3^{1/3}}
\label{garrphot}
\een
where $r_s = 2M/M_4^2$. However, it can
be seen from Eq.(\ref{garrphot}) that 
no real solution for $r_{\mathrm ph}$  exists for admissible values of
$l$ and $r_s$. This is to be expected since the metric
({\ref{gartanmetric}) represents a weak field solution.
The weak
field limit of the bending angle was obtained to be\cite{kar}
\be
\alpha_{\mathrm weak} = \frac{4M}{M_4^2 \tilde{r}_0} + \frac{4Ml^2}{M_4^2\tilde{r}_0^3}
\ee
where $\tilde{r}_0$ in this case is the isotropic coordinate equivalent of
the distance of closest approach $r_0$ in standard coordinates.

The analysis of strong field gravitational lensing can however be performed
in other braneworld geometries. For example,
let us consider lensing in the metric with tidal charge (\ref{tidalmetric})
given by\cite{dadhich}
\be
dS_4^2=-\left(1-\frac{2M}{M_4^2r}+\frac{Q}{r^2}\right)dt^2\\+
\left(1-\frac{2M}{M_4^2r}+\frac{Q}{r^2}\right)^{-1}dr^2\\ +r^2(d\Omega^2)
\label{tidalmetric-2}
\ee
which resembles the Reissner-Nordstrom metric, but with $Q < 0$ in the 
braneworld context. Again using units of distance $2M/M_4^2$, it is 
straightforward to obtain the expressions for the photon sphere and the
minimum impact parameter given by\cite{bozza}
\ben
r_{ph} &=& \frac{\left(3+\sqrt{9-32Q}\right)}{4} \\
u_{m} &=& \frac{\left(3+\sqrt{9-32Q}\right)^2}{4\sqrt{2}\sqrt{3-8Q+\sqrt{9-32Q}}}
\een
The coefficients $\overline{a}$ and $\overline{b}$ in the deflection angle
are given by\cite{bozza}
\ben
\overline{a} = \frac{r_{\mathrm ph}\sqrt{r_{\mathrm ph} - 2Q}}{\sqrt{(3 -
r_{\mathrm ph})r_{\mathrm ph}^2 - 9Q r_{\mathrm ph} + 8Q^2}}\\
\overline{b} = -\pi + 2 {\mathrm log}[6(2-\sqrt{3})] + \frac{8Q 
\Bigl(\sqrt{3} - 4 + {\mathrm log}[6(2-\sqrt{3})]\Bigr)}{9} \nonumber \\
+ \frac{(r_{\mathrm ph} - Q)^2 [(3- r_{\mathrm ph})r_{\mathrm ph}^2 -
9Q r_{\mathrm ph} + 8Q^2] \overline{a}{\mathrm log}[2]}{(r_{\mathrm ph}-
2Q)^3(r_{\mathrm ph}^2 - r_{\mathrm ph} + Q)}
\een 
In terms of the above coefficients one obtains the complete expression for
the deflection angle using Eq.(\ref{deflangle2}).
The weak field limit of the bending angle was derived to be\cite{kar}
\be
\alpha_{\mathrm weak} = \biggl(\frac{1}{M_4^2} - \frac{3\pi Q}{16M r_0}
\biggr)\frac{4M}{r_0}
\ee
Note that the bending angle is always positive because of negative tidal 
charge $Q$ unlike the electric charge of the Reissner-Noredstrom metric.

We next consider the braneworld solution (\ref{casad1}) in terms of the
PPN parameter $\beta$ given by\cite{casadio} 
\be
ds_4^2= -(1-\frac{2M}{M_4^2r})dt^2 + \frac{1-\frac{3M}{2M_4^2r}}
{(1-\frac{2M}{M_4^2r})
\left(1-\frac{M(4\beta-1)}{2M_4^2r}\right)} + r^2(d\Omega^2)
\label{casad1-2}
\ee
For the above metric the expressions for the radius of the photon sphere 
$r_{\mathrm ph}$ and the 
minimum impact parameter are of course similar to those for Schwarzschild
lensing given by Eqs.(\ref{phot}) and (\ref{minimp}), as is easy to see
using Eqs.(\ref{sphersym}), (\ref{photsphere}) and (\ref{impact}). The
expression for the deflection angle can be derived in terms of the 
coefficients $\overline{a}$ and $\overline{b}$ which for the metric
(\ref{casad1-2}) are given by (setting the unit of distance as $2M/M_4^2$)
\be
\overline{a} = \frac{\sqrt{3}}{\sqrt{6-(4\beta-1)}} 
\label{casadbend1}
\ee
\be
\overline{b} = -\pi + \frac{2\sqrt{3}}{\sqrt{6-(4\beta-1)}}{\mathrm log}[6(2-\sqrt{3})] +
\frac{\sqrt{3}}{\sqrt{6-(4\beta-1)}}{\mathrm log}[6]
\label{casadbend2}
\ee
Kar and Sinha\cite{kar} obtained the weak field limit of the bending
angle for the metric (\ref{casad1-2}) to be
\be
\alpha_{\mathrm weak} = \frac{2M(1+\beta)}{r_0}
\label{casadweak}
\ee
Note that the standard Schwarzschild expressions are recovered in 
Eqs.(\ref{casadbend1}), (\ref{casadbend2}) and (\ref{casadweak}), as should be,
for $\beta =1$. 

The lensing observables ${\cal S}$ and ${\cal R}$ 
corresponding to the ratio between the flux of the first
image and the flux coming from all the other images
as defined in Eqs.(\ref{obs-s}) and (\ref{obs-r}) respectively, can be
computed using the expressions for $u_m$, $\overline{a}$, and $\overline{b}$
for particular geometries. One can obtain the magnitudes of ${\cal S}$ and
${\cal R}$ for particular lensing candidates using the known values for their
masses and distances. For the black hole located at the centre of
our galaxy at a distance of $D_d = 8.5$kpc and with mass 
$M = 2.8 \times 10^6M_{\odot}$, 
the position of relativistic images
in Schwarzschild strong lensing was first computed by
Virbhadra and Ellis\cite{virbhadra1}, and the observable parameters ${\cal S}$
and ${\cal R}$ were computed by Bozza\cite{bozza}.
Whisker\cite{whisker} has computed using the above black hole as a candidate 
lense the values for these observables for two possible braneworld gemetries
given by the metric (\ref{tidalmetric-2}) with tidal charge\cite{dadhich}
and another solution\cite{gregory2}. 

\begin{table}[ph]
\tbl{Estimates for the strong field lensing angle
coefficients and observables for the black hole at the center of our galaxy 
with mass $M=2.8\times10^{6}M_{\odot}$ and $D_d = 8.5$ kpc, for standard
Schwarzschild and two different braneworld geometries. ($r_m = 2.5 
{\mathrm log}{\cal R}$.)}
{\begin{tabular}{|l|l|l|l|l|l|l|l|} 
\hline
{Observables}& {Schwarzschild} & \multicolumn{3}{|l|} {Brane metric} & \multicolumn{2}{|l|}
{Brane metric}\\
$\>$ &{metric} &\multicolumn{3}{|l|}{with tidal charge Q} & \multicolumn{2}{|l|}{with PPN parameter $\beta$}\\
\cline{3-7}
$\>$ & $\>$ & $Q=-0.1$ & $Q=-0.2$ & $Q=-0.3$ & $\beta=1+ 10^{-4}$ & $\beta=1- 10^{-4}$\\
\hline
$\theta_{\infty} (\mu\phantom{x}\textrm{arc sec})$&16.87 & 17.87 & 18.92 & 19.65 & 16.87 & 16.87\\
\hline
${\cal S}(\mu\phantom{x}\textrm{arc sec})$& 0.0211 & 0.0142 & 0.0102 & 0.097 & 0.02115
& 0.01923\\
\hline
$r_{m} (\textrm {magnitudes})$& 6.82  & 7.02 & 7.2 & 7.37 & 6.818 & 6.887\\
\hline
$u_m/r_s $& 2.6 & 2.75 & 2.9 & 3 & 2.6 & 2.6\\
\hline
$\overline{a}$& 1 & 0.9708 & 0.938 & 0.925 & 1.00006 & 0.9999\\
\hline
$\overline{b}$& -0.4002 & -0.612 & -0.747 & -0.819 & -0.402 & -0.429\\
\hline
\end{tabular} \label{ta1}}
\end{table}

In Table~\ref{ta1} we display the values of
these lensing quantities $\theta_{\infty}, {\cal S},  {\cal R}, u_m/r_s, 
\overline{a}$, and $\overline{b}$  first for Schwarzschild lensing. 
We compare the
values of the lensing observables with those obtained for lensing with
two braneworld geometries, i.e., the metric (\ref{tidalmetric-2}) with tidal 
charge\cite{dadhich}, and the metric (\ref{casad1-2}) with the PPN parameter
$\beta$\cite{casadio}. We choose three negative values of the tidal charge $Q$.
The two values of the parameter $\beta$ on either side of the Schwarzschild
value ($\beta = 1$) are chosen to maintain observational compatibility
with the solar system tests of the Nortvedt effect\cite{will}.
The measurement of the observable $\theta_{\infty}$ involves a microsecond
resolution which should in principle be attainable by the very long baseline
interferometry projects such
as MAXIM\cite{maxim}. Though the actual identification
of faint relativistic images would be extremely difficult in practice 
due to the inherent disturbances\cite{virbhadra1}, an accurate measurement
of $\theta_{\infty}$ would be able to distinguish the Schwarzschild
geometry from the braneworld RN-type one. However, to unambiguously determine
the exact nature of the black hole through the lensing angle coefficients
$\overline{a}$ and $\overline{b}$, one has to measure the observables
${\cal S}$ and ${\cal R}$. Since this involves the resolution of two faint
images separated by $\sim 0.02 \mu$ arc sec, such an observation would
need a leap of technological development over the present astronomical 
facilities\cite{whisker,bozza}.

\begin{figure}[pb]
%\begin{center}
\centerline{\psfig{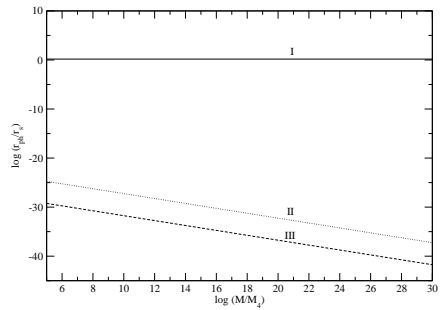}}
\vspace*{8pt}
\caption{The photon sphere $r_{\mathrm ph}$ (in units of the Schwarzschild
radius $r_s$) is plotted versus mass for the Schwarzschild black hole~(I),
and the Myers-Perry braneworld black holes with $l= 10^{30}l_4$~(II), and
$l= 10^{20}l_4$~(III).
\label{f2}}
%\end{center}
\end{figure}

Let us finally return to the Myers-Perry metric (\ref{smallmetric}) which is 
obtained from the $5$-dimensional Schwarzschild metric\cite{myers} and
could be relevant for the geometry near the horizon of a small 
$r \le l$ braneworld black hole. The cosmological evolution of such
black holes formed in the early universe having the metric
\be
dS_{4}^2=-\left(1-\frac{r_{BH}^2}{r^2}\right)dt^2+\left(1-\frac{r_{BH}^2}{r^2}\right)^{-1}dr^2\\+r^2\left(d\Omega^2\right)
\label{metric11}
\ee
has been described in details in section 4. It is possible for such black
holes to survive through the intermediate eras of the 
universe\cite{majumdar2,clancy}. It is feasible that such black
holes with masses in the sub-lunar range exist as dark matter in galactic 
haloes\cite{inoue,majumdar3}. The photon sphere for a Myers-Perry braneworld
black hole in units of $2M/M_4^2$ is given by
\be
r_{\mathrm ph} = \frac{2}{\sqrt{3\pi}}\sqrt{\frac{l}{l_4}}\sqrt{\frac{M_4}{M}}
\ee
and is plotted as a function of black hole mass in Figure~\ref{f2}.
The minimum impact parameter is
\be
u_m = \sqrt{2}r_{\mathrm ph}
\ee
The coefficients $\overline{a}$ and $\overline{b}$ given by\cite{eiroa}
\ben
\overline{a} = \frac{1}{\sqrt{2}} \\
\overline{b} = -\pi + \sqrt{2} {\mathrm log}(4\sqrt{2})
\een
can be then be used to construct the strong field deflection angle.
Eiroa\cite{eiroa} has calculated the positions and magnifications of the 
relativistic images and compared them with those of Schwarzschild black holes.

Gravitational lensing in the weak field limit by the black hole 
(\ref{metric11}) has been worked out\cite{majumdar4}. When the 
impact parameter $u$ exceeds a few times the horizon radius given by
Eq.(\ref{massradius}), but is still lesser than $l$ ($u \le l$), the 
application of weak field lensing could be relevant.
The weak field limit of the deflection angle was calculated to 
be\cite{majumdar4}
\be
\alpha_{\mathrm weak} = \frac{2Mll_4}{M_4r_0^2}
\ee
In order to satisfy the requirement that $u \le l$, and also obtain
non-negligible magnification at the image location, the mass of the black
hole should be such that\cite{majumdar4}
\be
\frac{M}{M_4} \le \frac{l}{l_4}
\label{masslim12}
\ee
Using the maximum allowed value for $l$ by present experiments\cite{long}, 
one then
obtains that $M \le 10^{-8}M_{\odot}$. So weak field gravitational lensing
for such a black hole could be applicable only for masses in the sub-lunar
range. As discussed above the braneworld scenario is conducive to the 
existence of primordial black holes in the sublunar mass 
range\cite{majumdar2,clancy}. If such 
black holes exist in our galactic halo, then the magnification of their
weak field images turns out to be diminished compared to the standard 
Schwarzschild black holes of similar mass\cite{majumdar4}.

\begin{table}[ph]
\tbl{Estimates for the strong field lensing angle
coefficients and observables for a black hole in the galactic halo
with mass $M=10^{25}$ gm and $D_d = 10^{22}$ cm, for the standard
Schwarzschild and the Myers-Perry geometry with two different values of $l$. 
($r_m = 2.5 
{\mathrm log}{\cal R}$.)}
{\begin{tabular}{|l|l|l|l|l|l|l|l|} 
\hline
{Observables}& \multicolumn{2}{|l|} {Myers-Perry metric} & {Schwarzschild}\\
$\>$ &\multicolumn{2}{|l|} {with extra dimention $l$}  &{metric}\\
\cline{2-3}
$\>$ & $l=10^{30}l_4$ & $l=10^{20}l_4$ & $\>$\\
\hline
$\theta_{\infty}(\mu\phantom{x}\textrm{arc sec})$&$0.039\times 10^{-12}$
&$0.0123\times10^{-16}$&$0.03\times 10^{-12}$\\
\hline
${\cal S}(\mu\phantom{x}\textrm{arc sec})$&$0.2042\times 10^{-17}$&$0.0644\times
10^{-21}$&$.0375\times 10^{-15}$\\
\hline
$r_{m} (\textrm {magnitudes})$&9.64&9.64&6.82\\
\hline
$u_m/r_s$&3.37&$1.065\times 10^{-4}$&2.6\\
\hline
$\overline{a}$&0.707&0.707&1\\
\hline
$\overline{b}$&-0.689&-0.689&-0.4002\\
\hline
\end{tabular} \label{ta2}}
\end{table}

Before concluding this section we present a comparitive study of lensing
in the Myers-Perry geometry (\ref{metric11}) with that in 
the standard Schwarzschild geometry. The
respective photon spheres are plotted as a function of mass in Figure~\ref{f2}
choosing two values of $l$ for the braneworld case. The mass range chosen is 
the one which could be relevant for Myers-Perry braneworld
black holes (\ref{metric11}) that have been conjectured to exist in
the form of binaries in the galactic halo\cite{majumdar3,inoue}.
Choosing a specific value of mass in the above range, the strong field lensing
quantities have been evaluated separately
for the above two metrics. The strong lensing angle coefficients 
$\overline{a}$ and $\overline{b}$, and the observables $\theta_{\infty}$,
${\cal R}$ and ${\cal S}$ are displayed in Table~\ref{ta2}. 
It is interesting to note that a larger value for the scale of the extra
dimension $l$ takes the values of the observables $\theta_{\infty}$ and 
${\cal S}$ for the braneworld case
closer to those of Schwarzschild lensing. This happens because the size of
the braneworld black hole which is much smaller compared to the Schwarzschild
black hole of same mass increases with $l$ for fixed 
mass~(\ref{massradius}). Of course,
the values of these lensing quantities are far beyond the possiblity 
of verification
by present observational capabilities. But the comparison of these numbers
provides an in principle method of discriminating between different
gravity models.

\section{Summary and Conclusions}

Braneworld black holes are the potential testing arenas of a rich 
theoretical structure associated with modified braneworld gravity and
extra dimensions. If the fundamental scale of gravity is much lower
than the $4$-dimensional Planck scale, then it is feasible for higher
dimensional black holes to be formed in low energy processes. 
A lot of the present interest in the prospect  of obtaining observable
signatures of  braneworld gravity is
via the properties of the evaporation products of mini black holes produced
either in particle collisions inside accelerators\cite{giddings,dimopoulos},
or in high energy cosmic ray showers\cite{feng}. Another issue of interest
is regarding  primordial
black holes which may be formed through the collapse of overdense regions
in the braneworld high energy phase of the early universe. Such black
holes  could have 
diverse cosmological implications. 
Further, even larger black holes could form by gravitational collapse
of matter on the brane. Braneworld black holes have entirely different
metrics compared to $4$-dimensional black holes, and carry the signature
of the extra dimension in their geometries. The focus of this review has
been to discuss the cosmological consequences of primordial braneworld 
black holes, and also to analyse the features associated with gravitational
lensing in several braneworld black hole metrics.

The physics of gravitational collapse on the brane is not yet understood
in totality\cite{germani}. A serious conceptual difficulty arises due to 
the fact that the
gravitational field equations are rendered incomplete on the brane by
the effect of the bulk gravitational modes. No complete solution to
the bulk gravitational field equations representing a spherically symmetric
vacuum black hole metric on the brane has been found till date. 
Projections of the $5$-dimensional Weyl tensor have been used in the
$4$-dimensional brane field equations in various configurations to
obtain possible braneworld black hole metrics.
In section~2 we have discussed some possible candidate geometries for
braneworld black holes. These metrics are in general rather different
from the standard Schwarzschild metric, and incorporate modifications of
the standard $1/r$ gravitational potential in braneworld gravity\cite{garrtan}.
The braneworld black hole solutions contain interesting features related
to the existence of horizons, and the modified black hole entropy and 
temperature.
The Reissner-Nordstrom type solution\cite{dadhich} contains
a negative tidal charge which can be viewed as the reflection of the black
hole mass by the bulk with negative cosmological constant. Other solutions
discussed include the one that incorporates the contribution of the projected
Weyl tensor in terms of the post-Newtonian parameters\cite{casadio}. 

The $5$-dimensional character of braneworld gravity at small scales
motivates the consideration of the induced Myers-Perry metric\cite{myers}
as a braneworld black hole candidate for black holes with radius $r < l$.
Physically this corresponds to the fact that a small black hole is unable
to distinguish between the bulk dimension and our $3$ brane dimensions.
The modified properties\cite{argyres} of such black holes within the 
braneworld context 
have been discussed\cite{emparan}, and the different subtelties regarding
their Hawking evaporation on the brane and into the bulk have been analysed
earler\cite{kanti}. Special features of the interaction of $5$-dimensional
black holes with the brane have been revealed in the 
literature\cite{frolov1,frolov3,frolov2,frolov4,frolov5}. In this review
we have highlighted
the relevance of using the Myers-Perry geometry for the analysis of
the processes of Hawking evaporation and the accretion of radiation 
in the early cosmological evolution of primordial black holes
in the braneworld scenario.

In section~3 we have provided a short summary of the main features of
the cosmology in the RS-II model\cite{randall2}. Essentially the cosmology
in the early braneworld era (high energy brane regime) is
altered by the presence of a term in the Friedmann equation on the brane
that is quadratic in the energy momentum tensor. This modifies the Hubble
expansion at times earlier than the time $t_c \equiv l/2$. Among other
consequences, such a scenario allows for the possibility of inflation
with steep potentials that could be excluded in the standard scenario.
Various features of the cosmology of braneworld models have been 
analysed in details in some recent reviews\cite{maartens,brax}.
In the context of the present review, the most significant implication of
the modified high energy behaviour is that the expansion of the Hubble volume
makes it feasible for accretion by the primordial black holes of the
surrounding radiation to take place. This feature is primarily responsible
for the  mass growth and prolonged survival\cite{majumdar2,clancy} of the 
braneworld black holes. The standard cosmological expansion of the universe
is recovered at later times, in order for the observationally established 
processes such as nucleosynthesis to work out. 

The description  of the braneworld cosmological evolution with primordial
black holes has been provided in section~4. The modified mass-radius
relationship for the induced $4$-dimensional Myers-Perry black holes leads
to slower evaporation and longer lifetimes compared to standard Schwarzschild
black holes\cite{guedens}. The detailed properties of evaporation of such
black holes are described in a recent review\cite{kanti}. The accretion of 
radiation in the high energy
phase could be effective because the growth of black hole mass is smaller
than the growth of the mass in the Hubble volume in braneworld 
evolution\cite{majumdar2,clancy}. Thus, a large fraction of primordial
black holes may survive up to much later eras. The decaying black holes
affect several cosmological processes at different eras, 
and hence the observational
abundance of different species, for example, the background high enery 
photons\cite{sendouda1},
could be used to put constraints on the initial mass spectrum of the
black holes. These contraints have to be evaluated considering the altered
cosmological evolution in the braneworld scenario, and therefore could be 
significantly modified compared to similar constraints in standard 
cosmology\cite{liddle}. The exchange of energy of the black holes with
the surrounding radiation in the high energy era leads to mass disequilibration
of neighbouring black holes\cite{majumdar3}. As a consequence, binaries could
be formed later through $3$-body gravitational interactions. Such binaries
have masses in the sub-lunar range, and gravitational waves emitted during
their coalescence come in the detectable range\cite{inoue}. 

Braneworld black holes exisiting at present times offer another scope of
detection, viz. through the gravitational lensing of light sources by
them. In section~5 we have first briefly reviewed the framework of 
gravitational lensing using which both weak field and strong field lensing
can be handled in a unified manner\cite{bozza}. The expressions for the 
lensing quantities in different
braneworld metrics have been presented analysing the crucial differences
from standard Schwarzschild lensing. Myers-Perry black holes existing in
the galactic halo in certain mass ranges would be difficult to detect via 
weak field lensing due to reduced magnification compared to Schwarzschild
black holes\cite{majumdar4}. Strong field gravitational lensing through its
prominent features such as the production of relativistic images and
retrolensing offers the ultimate scope of discriminating between different
gravity models\cite{eiroa}. Using the black hole at the centre of our galaxy
as a candidate lense, it is possible to compute the theoretical values of 
several lensing observables in different geometries\cite{whisker}.
We have presented the computed lensing observables for a few braneworld
metrics. A comparison with the lensing observables of the Schwarzschild
metric shows that it might be possible to distinguish the braneworld black hole
with tidal charge\cite{dadhich} from the Schwarzschild black hole by the
possible measurement of one of these observables in the near future.

The physics of extra dimensions in no longer a field of mere
theoretical constructs. 
There is currently a lot of ongoing activity on finding observational
signatures of braneworld black holes through different physical mechanisms.
In addition to the motion of light rays, the properties of massive orbiting
particles in these geometries are also expected to exhibit interesting 
features. In particular, it has been shown that no stable circular orbits 
exist in the equatorial plane of the Myers-Perry metric\cite{frolov5}.
The position of the innermost stable orbit shifts for braneworld black holes as
is the case with the photon sphere, and this could lead to observable 
modifications of the properties of the accretion disks around black holes.
Gravity wave spectroscopy offers another possibility of observing braneworld
gravity. It has been pointed out using the black string between two branes 
as a model of a braneworld black hole, that the massive bulk gravity modes
could lead to detectable spectroscopic signatures that are absent in
the normal $4$-dimensional gravitational waves\cite{seahra}.
Furthermore, the possibility of even distingushing between  different
braneworld models such as the ADD model\cite{arkani} and the 
RS model\cite{randall2} via the production of their respective black holes
in accelerators has been argued\cite{stojkovic2}. Much excitement exists
indeed in the prospects of detecting signatures of extra dimensions.
The range of cosmological and astrophysical implications of extra dimensions,
braneworld gravity, and black holes have only started to being investigated.

\end{document}